\documentclass[fleqn,11pt]{SelfArx} 

\usepackage{lipsum} 

\definecolor{color1}{RGB}{0,0,90} 
\definecolor{color2}{RGB}{0,20,20} 
\definecolor{color3}{RGB}{30,30,90} 

\usepackage{hyperref} 
\hypersetup{hidelinks,colorlinks,breaklinks=true,urlcolor=color2,citecolor=color1,linkcolor=color1,bookmarksopen=false,pdftitle={Title},pdfauthor={Author}}

\JournalInfo{This is a preprint version of} 
\Archive{Trends in Microbiology, Vol. 23, No. 1, 4-6, 2015} 

\PaperTitle{Cell-size maintenance: universal strategy revealed} 

\Authors{Suckjoon Jun\textsuperscript{1,2}*, Sattar Taheri-Araghi\textsuperscript{1}} 
\affiliation{\textsuperscript{1}\textit{Department of Physics, University of California San Diego, La Jolla, California 92093, USA}} 
\affiliation{\textsuperscript{2}\textit{Section of Molecular Biology, Division of Biology, University of California San Diego, La Jolla, California 92093, USA}} 
\affiliation{*\textbf{Corresponding author}: S. Jun, suckjoon.jun@gmail.com} 

\newcommand{\keywordname}{Keywords} 

\Abstract{How cells maintain a stable size has fascinated scientists since the beginning of modern biology, but has remained largely mysterious. Recently, however, the ability to analyze single bacteria in real time has provided new, important quantitative insights into this long-standing question in cell biology.}

\begin{document}

\flushbottom 

\maketitle 


\thispagestyle{empty} 


In nature, cells can be as small as $\sim$0.2 mm (e.g., \textit{Mycoplasma gallicepticum}) and as large as $\sim$0.1 m (e.g., \textit{Syringammina fragilissima}), spanning almost six orders of magnitude. Individual organisms, however, show much narrower size distributions, and under constant conditions most single-celled microorganisms change their size by only two-fold between birth and division. For \textit{Escherichia coli}, the variance of size distribution at division is $\sim$10\% of the average~\cite{1}, a strong indication that these cells know how to maintain stable size.

In 1958, Schaechter, Maal{\o}e, and Kjeldgaard established a general underlying principle in microbial physiology known as the `growth law'~\cite{2}. It states that the average cell size is exponentially proportional to the average nutrient-imposed growth rate. That is, if we culture the cells in an unknown liquid medium X, we just need to measure the growth curve and we can predict the exact average size of the cells in that medium. What determines the cell size, and how do cells maintain their size under a given growth condition?

\begin{figure*}[!t]
\centering
\includegraphics[width=1\textwidth]{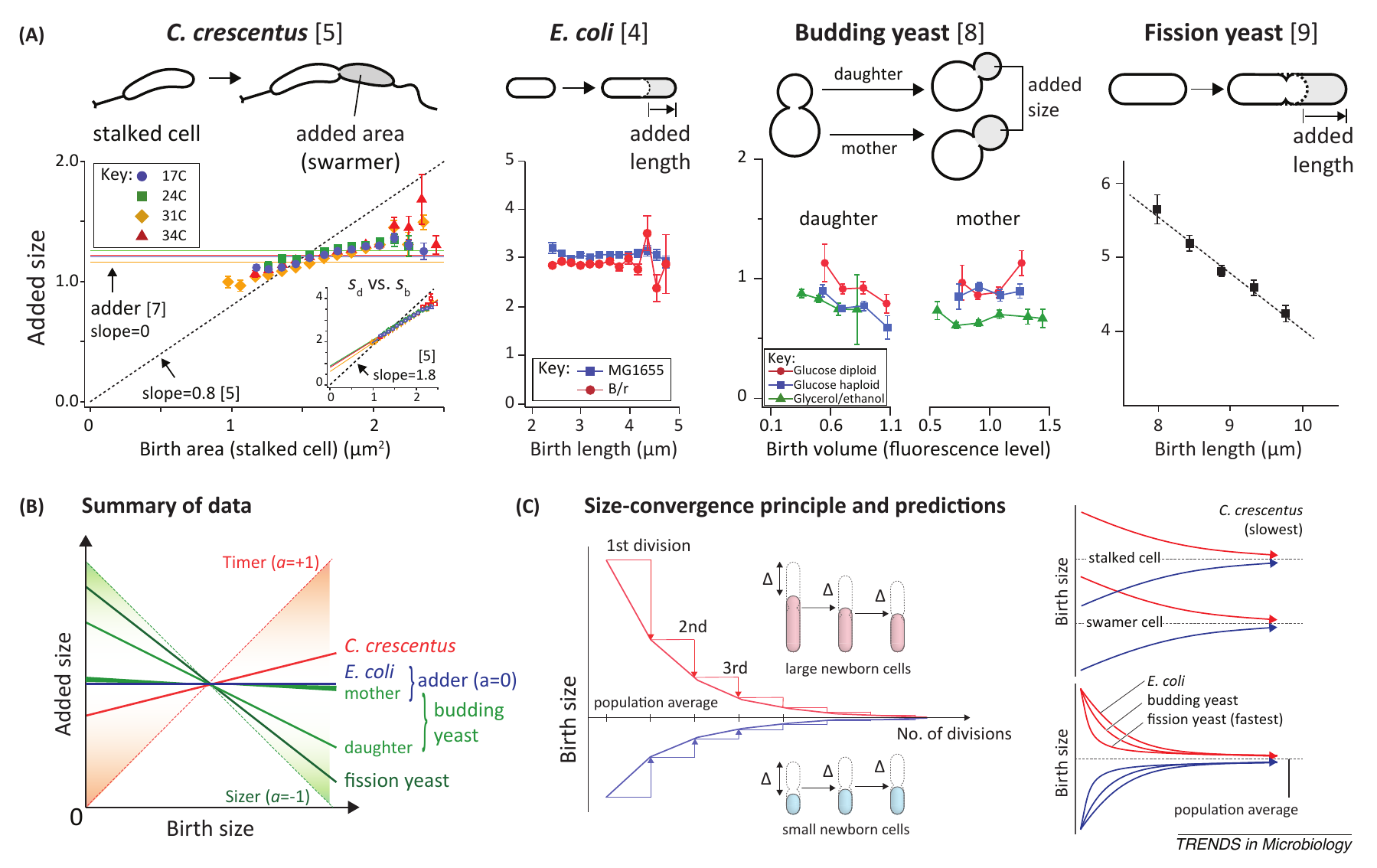}
\caption{\small
Relationship between size at birth ($s_b$) and added size between birth and division reveals the nature of cell-size maintenance. (A) The published data of added size from bacteria to single-celled eukaryotes is well described by a linear line $a.s_b + \Delta$, where a is the slope and $\Delta$ is the off-set. \textit{Caulobactor crescentus} shows weak positive slopes ($a = +0.21 \sim +0.33$)~\cite{5}, whereas the timer in~\cite{5} predicts $a = +0.8$. The inset is division size ($s_d$) vs. birth size ($s_b$), showing an unambiguous deviation from the timer prediction. This resembles the \textit{Escherichia coli} data~\cite{4}, where $a\approx0$ (perfect adder). \textit{E. coli} thus add constant size $\Delta$ irrespective of the newborn size. Budding yeast shows a range of slopes (from near $0$ for mother cells up to $-0.52$ for daughter cells; data from~\cite{8}), and fission yeast shows a strong negative slope ($a = -0.76$; data from~\cite{9}, adapted, with permission, from Journal of Cell Science). (B) The data in (A) can be summarized in a single added size vs. birth size diagram. Slope $+1$ corresponds to perfect timer, $-1$ to perfect sizer, and $0$ to perfect adder. Based on this criterion, \textit{C. crescents} is timer-like adder and yeast is sizer-like adder, whereas \textit{E. coli} is perfect adder. (C) It is thus unlikely that these model organisms share the same biological mechanism for size control. However, their size homeostasis can be described by the same principle illustrated here. For perfect adder, both large and small newborn cells passively approach the population average size by adding constant $\Delta$. This applies not only to imperfect adders such as yeast, but also to asymmetric dividers such as \textit{C. crescentus}. The major difference between these organisms is their speed of size convergence, and shown on the right is a schematic prediction based on the sign of the slope $a$.
}
\end{figure*}

Historically, cell size homeostasis has been discussed in the context of two major paradigms: sizer, in which the cell actively monitors its size and triggers the cell cycle once it reaches a critical size, and timer, in which the cell attempts to grow for a specific amount of time before division. 
Pinning down which model is correct poses daunting experimental challenges, because size control study requires quantitative measurements at the single-cell level with extreme precision~\cite{3} and throughput~\cite{4} under tightly controlled experimental conditions. It has only been in the past few years that the data with sufficient quantity and quality~\cite{4,5} have become available to address size maintenance in the way researchers since the 1950s dreamed of.

The latest in the series of single-cell studies is the collaborative work by the Scherer and Dinner groups~\cite{5}. They studied \textit{Caulobacter crescentus}, a model bacterial organism known for asymmetric cell division and cellular differentiation. 
Upon division, the two daughter cells of \textit{C. crescentus} are distinct from each other in shape and size: the larger `stalked' cell binds to a surface, whereas the smaller `swarmer' cell is initially motile and differentiates into a stalked cell. 
This allowed the authors long-term continuous observations of growth and division of the stalked cells in a flow chamber, producing amounts of single-cell data comparable to previous work in \textit{E. coli}~\cite{4}. 
The authors addressed two questions: 
($i$) How do cell size and generation time change with respect to the temperature-imposed growth rate? 
($ii$) What is the relationship between the size at birth and division?

The answer to the first question has been suspected by bacterial physiologists since the 1950s~\cite{2,6}. 
That is, temperature affects only the overall chemical reaction rates, and changing it is equivalent to rescaling the global timescale of physiology. 
Thus, watching growth and division of individual cells at different temperatures would be like playing the same film at different speeds. 
The large amounts of data in~\cite{5} indeed elegantly shows that the size and generation time distributions obtained at different temperatures collapse when rescaled by their respective means.

The second question is related to size maintenance. 
The authors plotted division size ($s_d$) vs. birth size ($s_b$) from individual cells, and concluded that the data are scattered around a linear line $s_d = 1.8 . s_b$. 
This is reminiscent of timer, because it would mean that cells 'divide upon reaching a critical multiple (1.8) of their initial sizes' after growing a specific amount of time~\cite{5}.

However, the conundrum is that timer cannot maintain stable size distributions when cells elongate exponentially, since size fluctuations diverge as a square root of the number of consecutive cell divisions, like a random walk (to prevent an uncontrolled size divergence, the birth size and the generation time should be negatively correlated~\cite{5}). 
In fact, close inspection of the data~\cite{5} (http://dinnergroup.uchicago.edu/downloads.html) suggests that \textit{C. crescentus} maintains stable size following a principle for \textit{E. coli} originally proposed by Koppes and colleagues. 
This principle states that cells add a constant size between birth and division, irrespective of the birth size~\cite{7}. 
The published data in \textit{E. coli}~\cite{4} strongly supports the model (Figure 1A and 1B; http://jun.ucsd.edu/mother\_machine.php). 
The beauty of this `adder' is that it automatically ensures size homeostasis, because at every cell division the cell approaches (albeit passively) the population average as illustrated in Figure 1C.

Size homeostasis requires neither perfect adder, nor symmetric division. 
For example, \textit{C. crescentus} actually shows weak positive slopes ($+0.21\sim+0.33$; data from~\cite{5}) between added size vs. newborn size, whereas budding yeast and fission yeast show negative slopes (up to $-0.8$; data from~\cite{8,9}). 
Yet, their size maintenance shares the same convergence principle for perfect adder (slope $0$; \textit{E. coli}) (Figure 1C).

[For interested readers, here is a fun exercise: Consider a newborn cell growing by $ a . s_b + \Delta$ as \textit{imperfect adder} (Figure 1B), where $a$ is the slope and $\Delta$ is the y-intercept (perfect adder means $a = 0$, i.e., growth by a constant size $\Delta$). One can show that the newborn size converges to $s_b\rightarrow\Delta/(1-a)$ for symmetric division, as long as $-1 < a < 1$ and $\Delta > 0$. For asymmetric dividers such as \textit{C. crescentus}, where the size ratio between daughter 1 and 2 is $r_1: r_2=(1-r_1)$ and daughter cell $1$ grows by $a. s_{1,b} + \Delta$, their newborn sizes converge to $s_{1,b} \rightarrow \Delta.  r_1/(1-(1+a). r_1)$ and $s_{2,b} \rightarrow \Delta . r_2/((1+a). r_2 - a)$. Thus, for perfect adder ($a = 0$) with symmetric division $r_1=r_2=1/2$, we recover $s_{1,b}\rightarrow \Delta$ \& $s_{2,b} \rightarrow \Delta$. For timer, $(1+a). r_1= 1$ and, therefore, $s_d = s_{1,b}/r_1 = (1 + a) . s_{1,b}$. If \textit{C. crescentus} employed timer as $s_d = 1.8. s_{1,b}$, the slope would be $a = 0.8$ (Figure 1A)].

The major difference between these organisms is their speed of size convergence. Quantitative prediction is straightforward, and adders with positive slopes (sizerlike) correct their size deviations faster than adders with negative slopes (timer-like) (Figure 1C).

After all, probably neither perfect timer ($a=+1$) nor perfect sizer ($a=-1$) exists in nature, and our simple criterion based on adder provides a general framework for understanding the nature of size maintenance. 
The next question is: What is the biological origin of the sign of the slope, and how is it related to the cell cycle to ensure one-to-one correspondence between replication and cell division~\cite{10}? Experimental methods that can probe physiology at the single-cell level, combined with quantitative analysis that can make experimentally testable predictions, will be key to answering these fundamental questions.

\section*{Acknowledgments} 
We thank Serena Bradde, Petra Levin, Massimo Vergassola for numerous discussions, Stefano Di Talia for his insight and the yeast data [8], Srividya Iyer-Biswas, Charles Wright, Aaron Dinner, and Norbert Scherer for the \textit{C. crescentus} data~\cite{5}, and Jeff Gore for suggesting the notion `adder'. S.J. is supported by the Paul G. Allen Foundation, the Pew Charitable Trusts, and National Science Foundation CAREER Award.

\bibliographystyle{unsrt}


\end{document}